\documentclass[summary]{ursi}

\title{Super-resolution of Ray-tracing Channel Simulation via Attention Mechanism based Deep Learning Model}

\author{Haoyang Zhang\affref{ref1}, Danping He \affref{ref1}\affref{ref2}, Xiping Wang\affref{ref1}, Wenbin Wang\affref{ref1}, Yunhao Cheng\affref{ref1}, Ke Guan\affref{ref1}}

\affiliation{%
  \aff{ref1}{State Key Laboratory of Rail Traffic Control and Safety, Beijing Jiaotong University, 100044, Beijing, China;}
  \aff{ref2}{Collaborative Innovation Center of Railway Traffic Safety, 100044, Beijing, China;}
  *Corresponding Author: Danping He (e-mail: hedanping@bjtu.edu.cn)
}

\begin{document}

\maketitle

\begin{abstract}
  As an emerging approach, deep learning plays an increasingly influential role in channel modeling. Traditional ray tracing (RT) methods of channel modeling tend to be inefficient and expensive. In this paper, we present a super-resolution (SR) model for channel characteristics. Residual connection and attention mechanism are applied to this convolutional neural network (CNN) model. Experiments prove that the proposed model can reduce the noise interference generated in the SR process and solve the problem of low efficiency of RT. The mean absolute error of our channel SR model on the PL achieves the effect of 2.82 dB with scale factor 2, the same accuracy as RT took only 52\% of the time in theory. Compared with vision transformer (ViT), the proposed model also demonstrates less running time and computing cost in SR of channel characteristics.  
\end{abstract}

\section{Introduction}
With the rapid development of wireless communication technology, 5G is widely used in various fields and daily life applications, such as media streaming, gaming, and video conferencing \cite{rZTE}. Accurate channel model is regarded as the foundation of future wireless network.
Generally, deterministic and semi-deterministic modeling are the two mainstream methods of channel modeling \cite{r3}. However, this method has many limitations, including low computational efficiency, excessive computing power consumption, and it requires complicated simulation.    

Many researchers are trying to make breakthroughs in channel modeling by machine learning (ML) \cite{RN58}. Because of its generalizable architecture, machine learning is widely used in almost every branch of science and technology. Channel modeling is not an exception \cite{Lina2}. As for ML, deep learning (DL) models of super-resolution (SR) such as CNN \cite{r16}, Transformer \cite{r17} and generative adversarial network (GAN) \cite{r18} are frequently employed. However, most studies are based on pure electromagnetic environment data without considering complex terrain and building distribution \cite{BJTU2}.

\begin{figure}[t]
\centering
\noindent
  \includegraphics[width=3.4in]{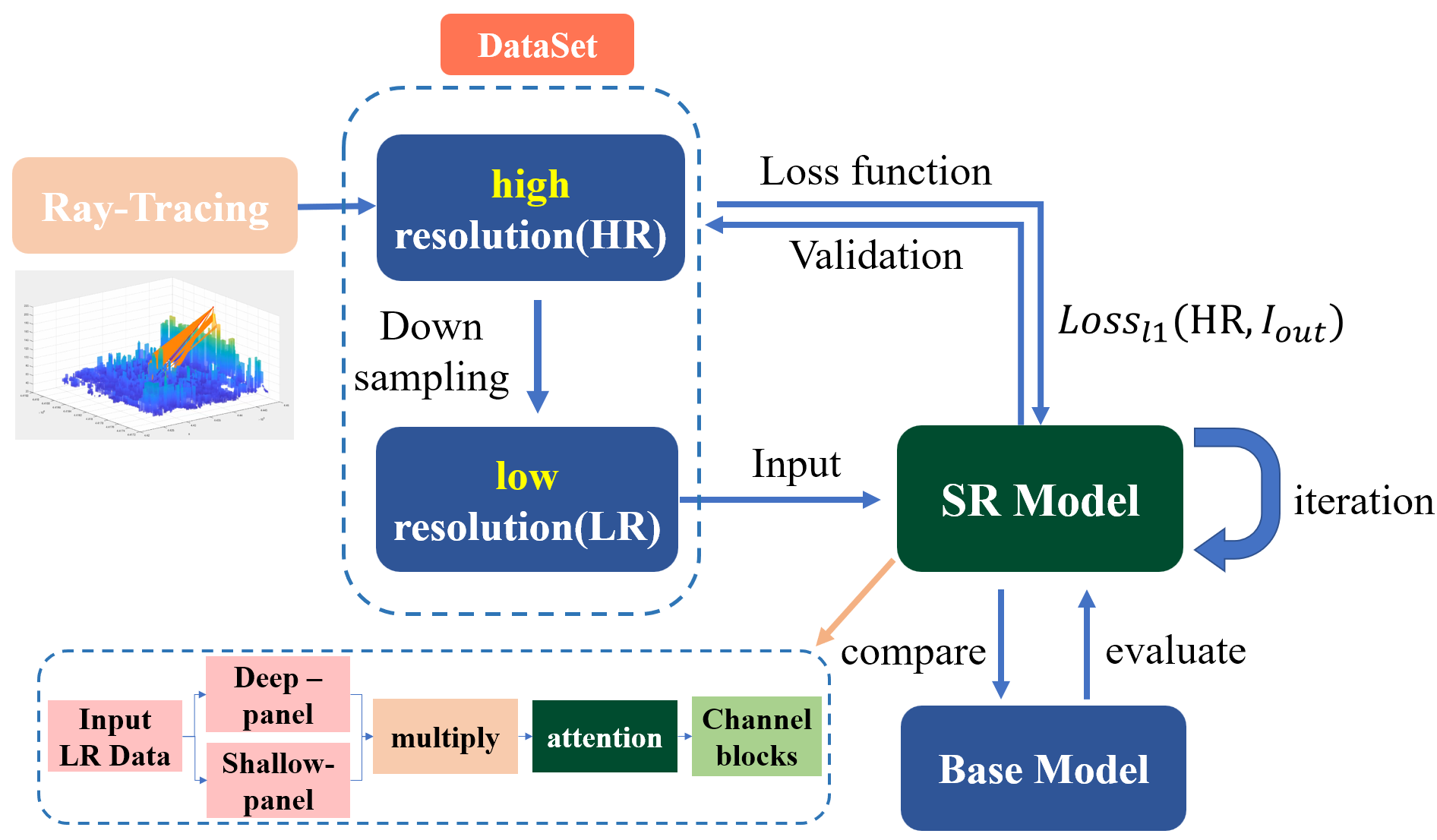}
  \caption{The overview of proposed SR model for wireless channel characteristics. }\label{overview}
\centering
\end{figure}

In this paper, we propose a DL model of SR for channel characteristics. It recovers high-resolution (HR) characteristics over low-resolution (LR) scenes built from original data. By using CNN with residual connection and attention mechanisms, we construct the proposed model. The overview of our model is shown in Fig. \ref{overview}. CloudRT \cite{r101} is used for RT simulation, simulating the propagation of radio waves in complex environments in dense urban areas, and obtaining channel characteristic information. Besides mean absolute error (MAE), we also incorporate the root mean square error (RMSE) and standard deviation error (STDE) into the loss function to balance reliability and error. We evaluate our proposed model by ablation study and comparisons with other SR models.
\section{Methodology}
\subsection{Data preprocessing and construction}
Utilizing CloudRT, we developed a high-precision channel feature dataset, \textbf{RT-urban},  which is used for training our proposed SR model. Simulation configuration details are summarized in \cite{r101}. Seven channel characteristics are obtained by simulation. The height of buildings (h), path loss (PL), multipath power ratio ($\textit{R}_p$), and LOS/NLOS area classification, root mean square (RMS) delay (DS), RMS azimuth angle spread ($\phi$), and RMS elevation angle spread ($\theta$). The latter six characteristics are our SR targets. 462 ray data were generated in more than 70 dense urban areas.

Values of channel characteristics that are far beyond ordinary thresholds in communication systems are set as the minimum (PL, $\textit{R}_p$) or the maximum (DS) of the corresponding normal range. NaN value represents the data of receivers located inside buildings. This value should be void but set as a real number out of the normal range so that the DL model can distinguish. We also use data augmentation. Through rotating the data map, the size of RT-urban is expanded to 6 times the original size. The related data and processing methods are shown in our previous work \cite{r101}.
\subsection{Neural network architecture}
The generation problem of the channel feature can be expressed as the SR problem of the image. The key to image SR lies in recovering from LR data to HR data. \cite{r114514}
\subsubsection{Deep-shallow pipe based on residual network}
In this study, we want to directionally generate various HR channel features from the collected and processed LR channel data. There are differences between different channel characters, and if all characters are fed into our network, the results will be unsatisfactory. So a method of classifying the propagation of deep-shallow channels is proposed.

\begin{figure}[t]
\centering
\noindent
  \includegraphics[width=3.4in]{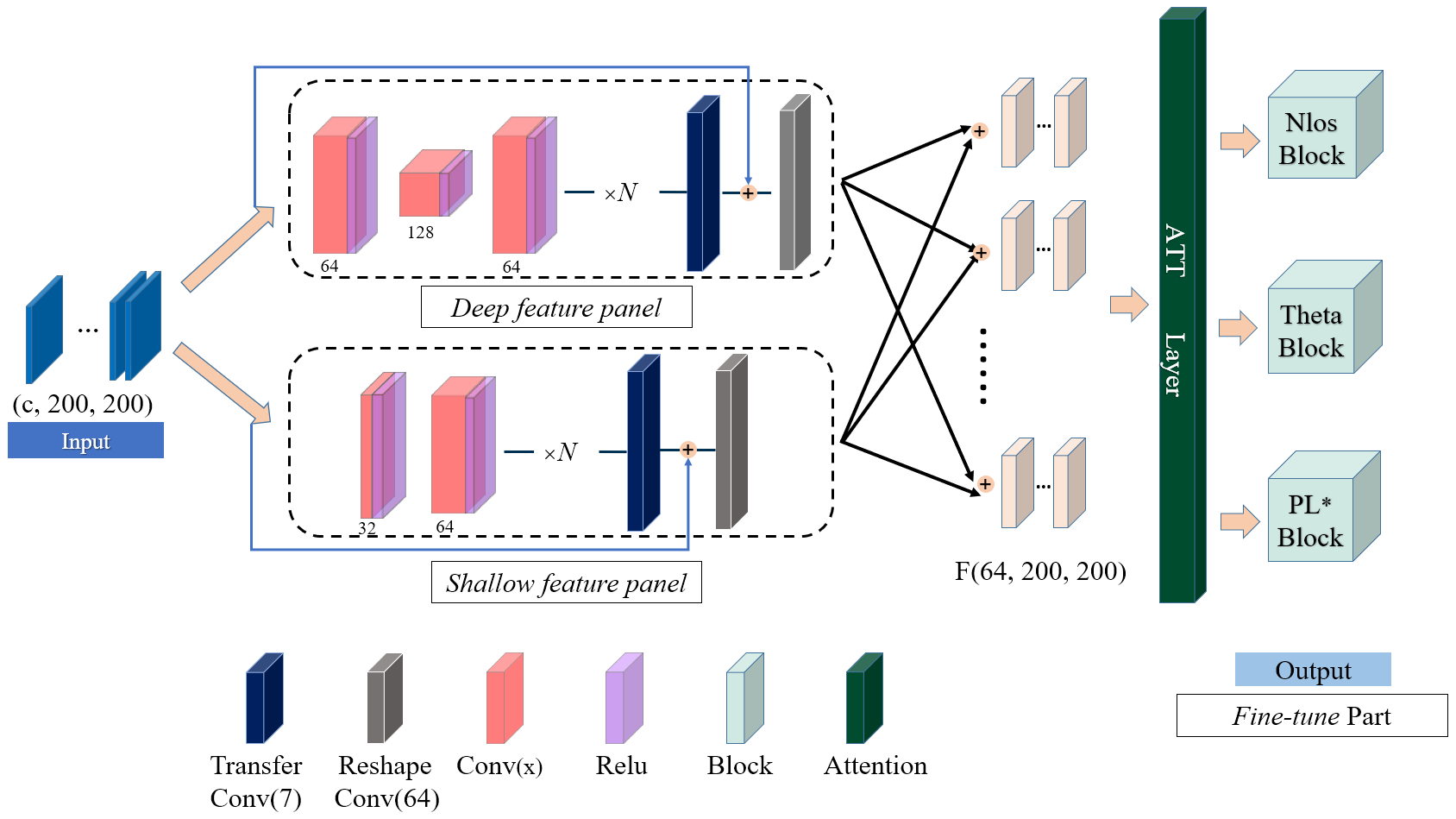}
  \caption{The overview of the proposed SR model with an attention mechanism. Conv(x) means that the convolution scale is x. PL* stands for PL, DS, and $R_{p}$.}\label{network}
\centering
\end{figure}

In Fig. \ref{network}, the deep-shallow backbone has a deep and shallow panel, which extracts features of different dimensions from input data. The deep panel has two convolutional layers with activation function ReLU more than the shallow panel and repeats the illustrated convolution block multiple times to extract features fully. The number of this block N in our work is 2. The method of residual connection is also introduced. It helps to expand the receptive field and reduce the loss of features caused by excessive convolution. 
\begin{equation}
\left\{
\begin{array}{lr}
H_{deep} = \mathcal{F}_{deep}(I_{in}, W_{d}) + I_{in}\\
H_{shallow} = \mathcal{F}_{shallow}(I_{in}, W_{s}) + I_{in}
\end{array}
\right.
\label{ILR}
\end{equation}
Where $H$ is the output after convolution, $I_{in}$ and $W$ present the input data and weights learn from deep and shallow panels $\mathcal{F}$. After each batch of convolution block, the unprocessed input data is residually connected to the feature maps produced by the operation. It can help prevent gradient dissipation during back-propagation, thus making it easier to train deep networks.
\subsubsection{Attention mechanisms and feature blocks}
\begin{figure}[h]
\begin{center}
\noindent
  \includegraphics[width=3.4in]{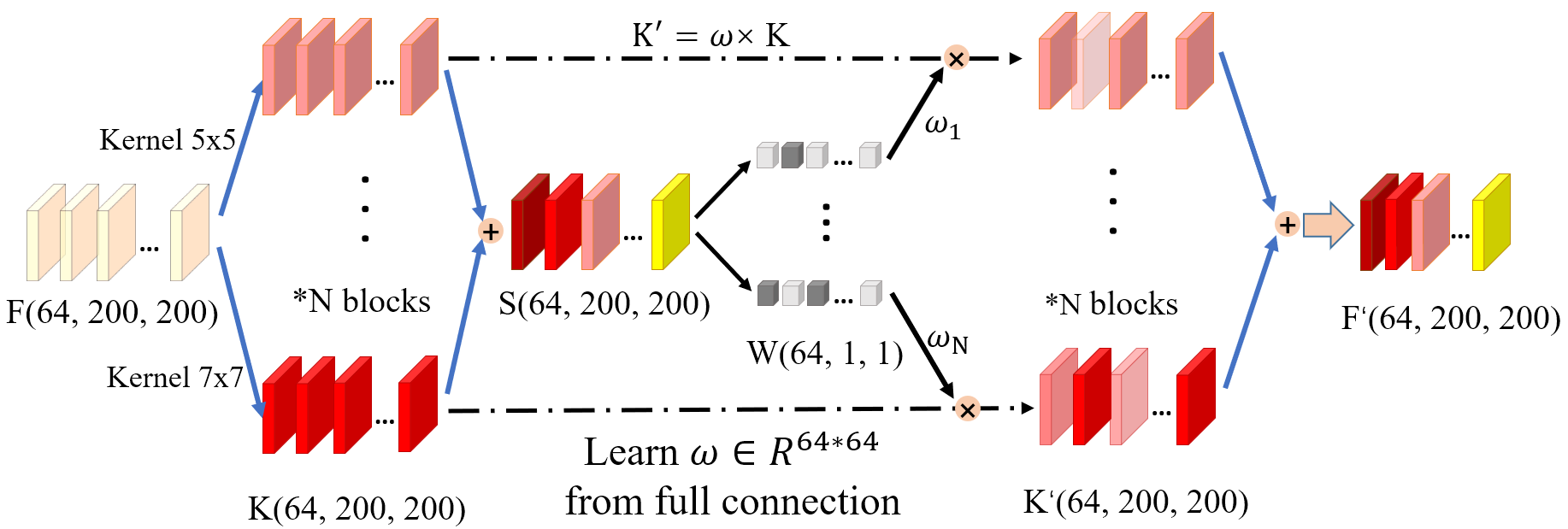}
  \caption{The overview of attention layer}\label{attention}
\end{center}
\end{figure}

It has to be considered that multi-path channel characteristics such as PL, DL, and LOS. The learning patterns of these features are completely different and tend to show distinguished differences in the learning process. So we introduced an attention mechanism for multi-feature extraction on the framework of the SR model in Fig. \ref{attention}. The feature map $F$ is split into several branches $K_{list}$ by using kernels with different sizes: 
\begin{equation}
K_{list}=Conv(F, kernel _{list}=[1, 3, 5, 7]_{size})
\label{ILR}
\end{equation}

As stated before, our goal is to control the information flows from multiple characteristics carrying different scales of information into the next layer. To achieve this, we need to integrate channels from the first N branches of the $K$ list:
\begin{equation}
    S = \sum_{i}K_{i},\quad i = [1,3,5,7] 
    \label{S}
\end{equation}
then we embed the global information by simply using average pooling to generate channels as $W \in \mathbb{R}^{C}$. The $C$ channels are calculated by the following in dimensions $h\times w$:
\begin{equation}
    W_{c} = \mathcal{F}_{avg}(S) = \frac{1}{h\times w}\sum^h_{i}\sum^w_{j}S(i,j)
    \label{weight}
\end{equation}
So it can get a compact weight $w\in \mathbb{R}^{64\times 64}$ of different channels by a fully connected layer, with the reduction of dimensionality for better efficiency. By fusing branches of features, this mechanism enhances the ability of the previous deep and shallow panels to extract characteristics and it will also pay attention to multi-scales of information when dell with different learning patterns of characteristics.
\subsubsection{Loss functions}
The proposed model use pixel loss instead of content loss for characteristics learning tasks. Through experiments, we found that the L1loss can help achieve better results on our task than the peak signal-to-noise ratio. In data preparation, we mentioned six different channel features in this task. However, the characteristic of LOS is unique. It only has two types of integer values. Therefore, the performance can only be judged by valuing the accuracy of the model for these two types of numerical classification. As a result, cross-entropy is used as the evaluation index.
\begin{equation}
loss_{l\small{1}}(\hat{I}, I) = \frac{n}{(hw)^2} \sum_{i, j}|\hat{I_{i, j}} - I_{i, j}|  \\
\label{ILR}
\end{equation}
\begin{equation}
loss_{ce}(\hat{I}, I) = -\frac{n}{(hw)^2} \sum_{i, j}I_{i,j,k}log\hat{I}_{i,j,k} \\
\label{ILR}
\end{equation}
Where $n,h,w$ represents the number of batches to be estimated and the length and width of the region we chose for input data, respectively.
In order to enhance the confidence of the fitted data, we also include the STDE as part of the evaluation index of the fitting effect of the loss function:
\begin{equation}
STDE(\hat{I},I) = \sqrt{\frac{1}{hw} \sum_{i,j}(\hat{I}_{i,j}-I_{i,j})^2}
\label{ILR}
\end{equation}

The standard deviation reflects the degree of dispersion between the pixel value of the image and the mean value. 

\section{Experiment}
\subsection{Performance of the proposed model}
\begin{figure}[ht]
	\centering
	%\subfigbottomskip=0.5pt %设置第二行子图与第一行子图的距离，即下面的头与上面的脚的距离
	\subfigure[MAE of path loss]{
		\label{PL_img}
		\includegraphics[width=0.48\linewidth]{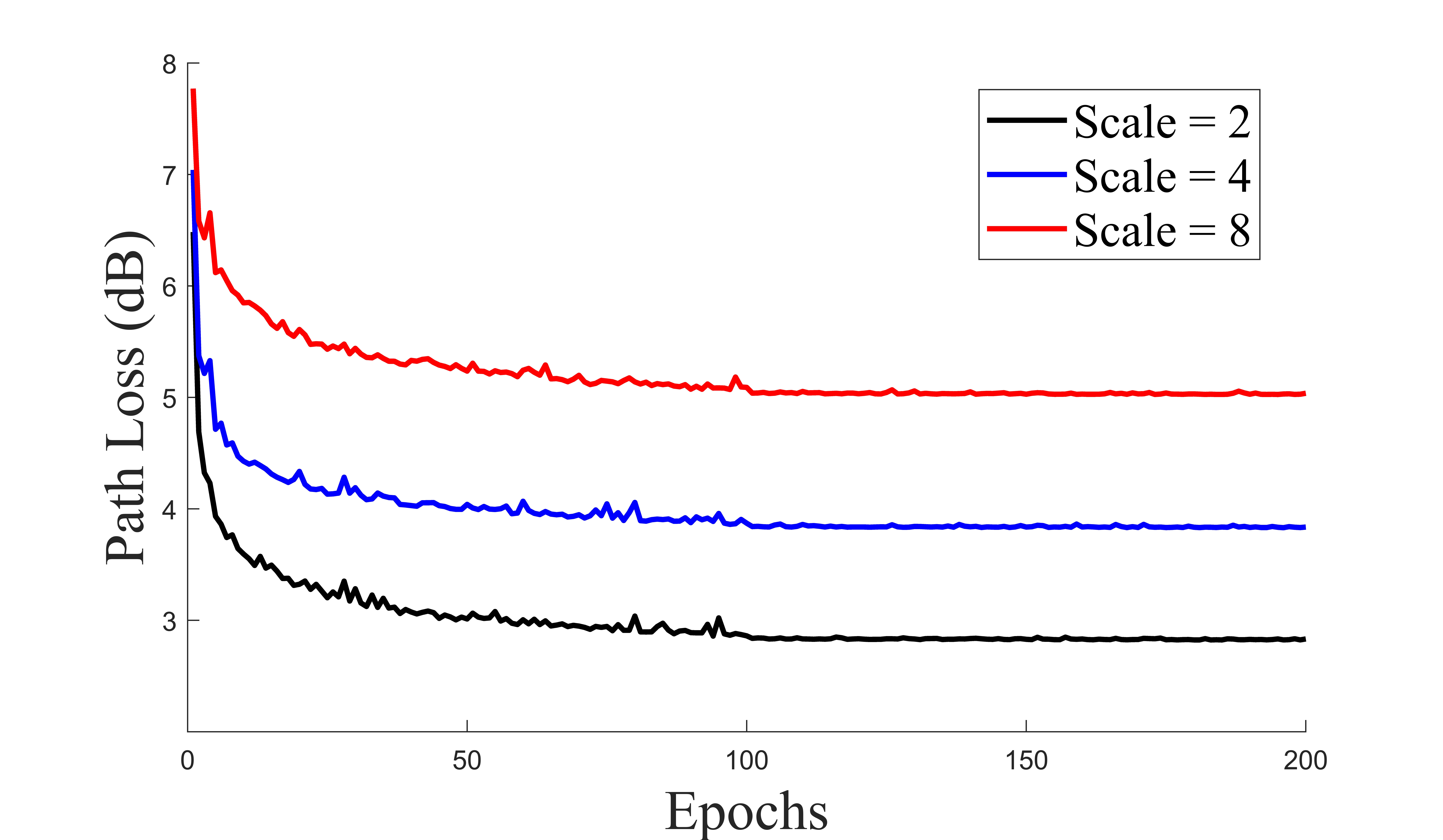}}
	\subfigure[Classification accuracy of LOS/NLOS]{
		\label{LOS_img}
		\includegraphics[width=0.48\linewidth]{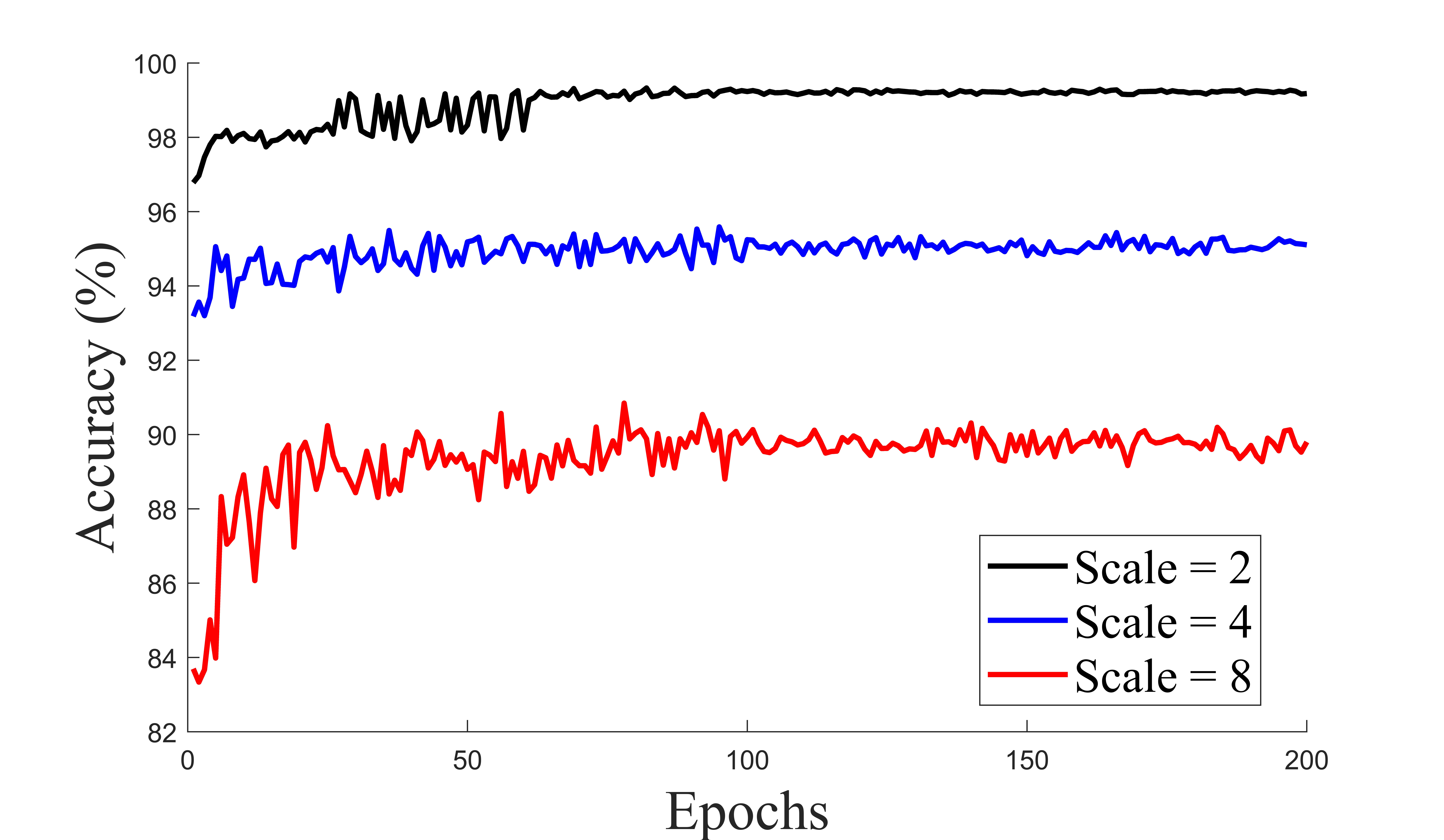}}
	\caption{MAE and classification accuracy of 2 main SR targets during the training process.}
	\label{trainprocess}
\end{figure}
\begin{figure}[ht]
	\centering
	\subfigbottomskip=0.5pt %设置第二行子图与第一行子图的距离，即下面的头与上面的脚的距离
	\subfigure[Ground-Truth]{
		\label{Ground_Truth}
		\includegraphics[width=0.31\linewidth]{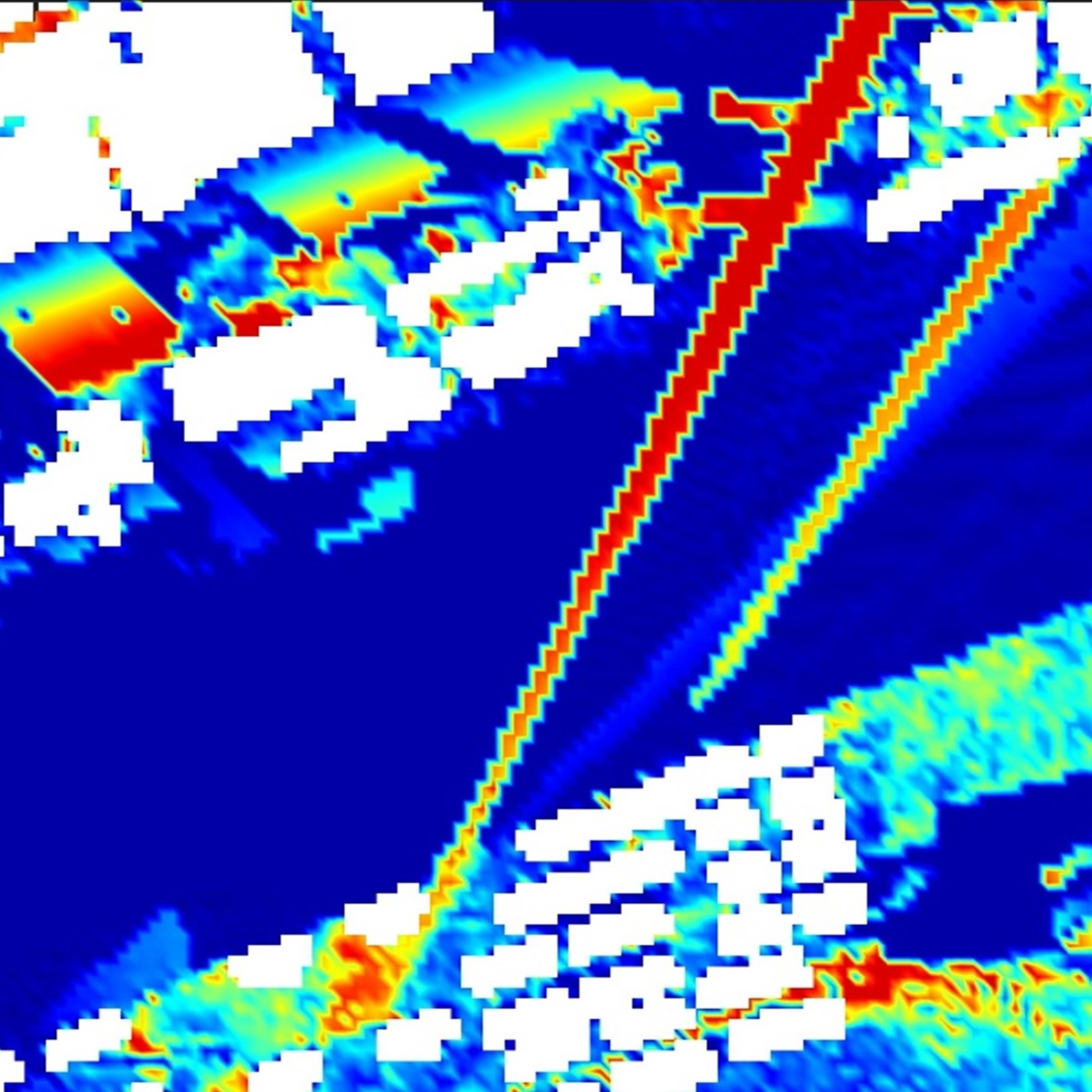}}
	\subfigure[UNet]{
		\label{Unet}
		\includegraphics[width=0.31\linewidth]{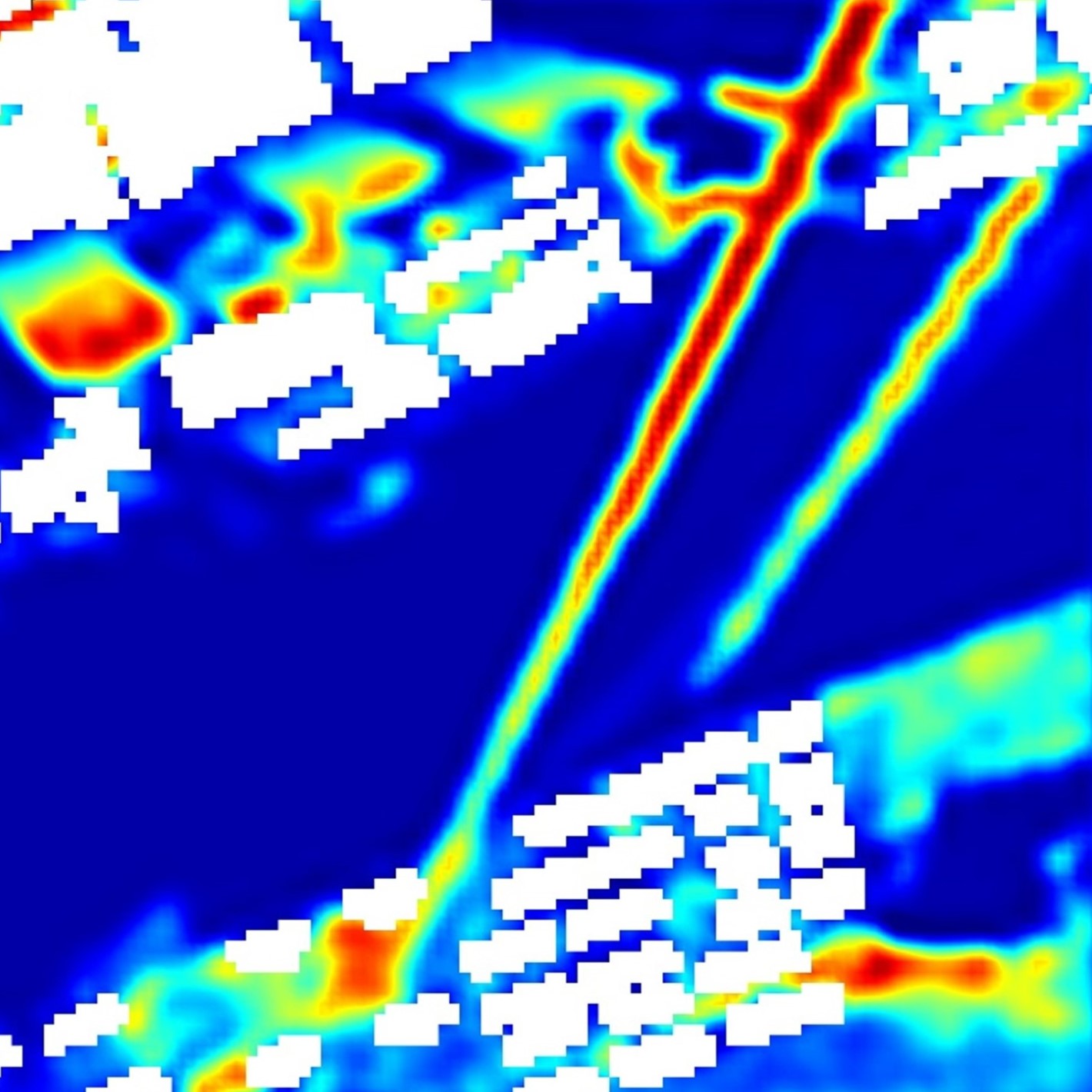}}
	\subfigure[ViT]{
		\label{ViT}
		\includegraphics[width=0.31\linewidth]{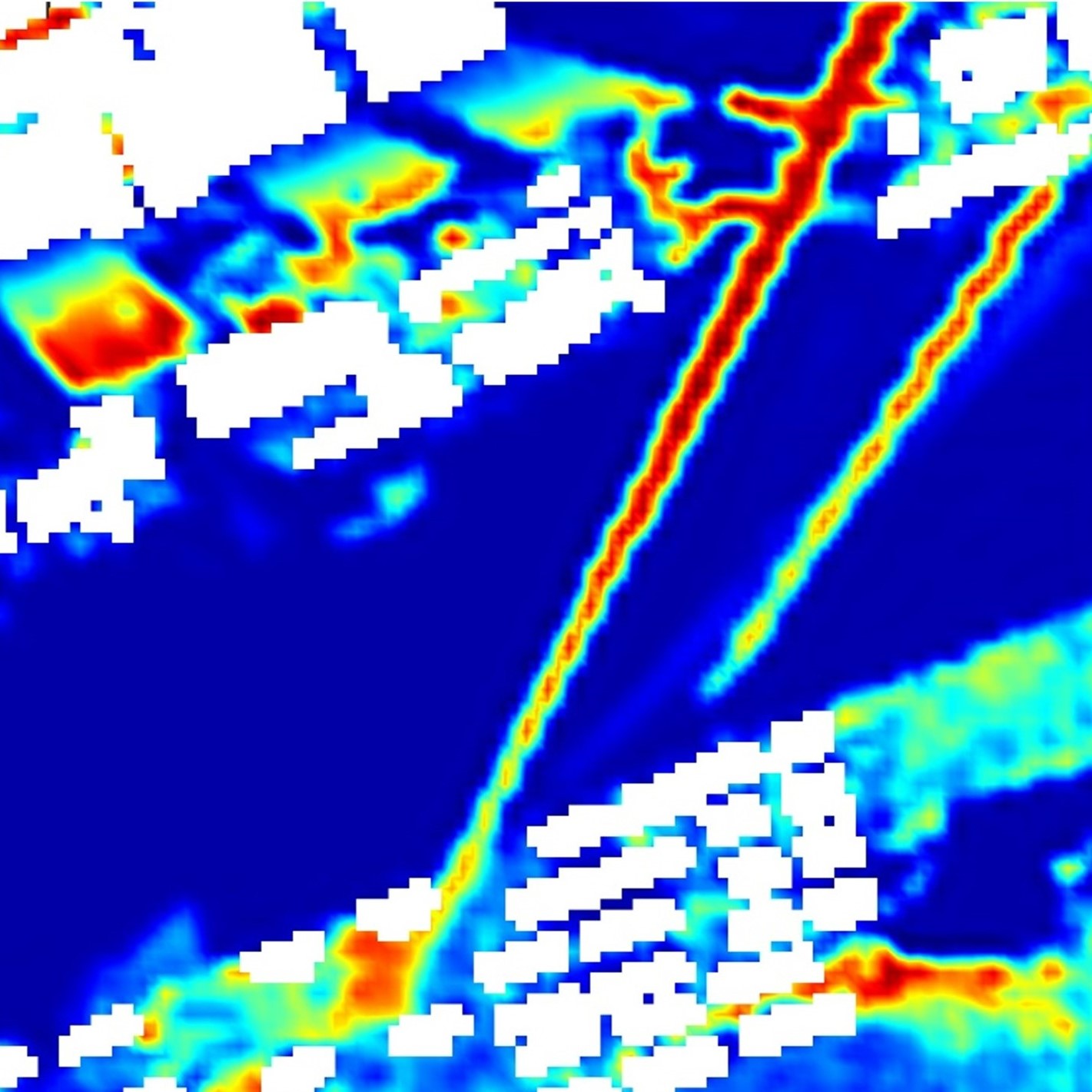}}

	%\qquad
	%让图片换行，
	
	\subfigure[Ours]{
		\label{Ours}
		\includegraphics[width=0.31\linewidth]{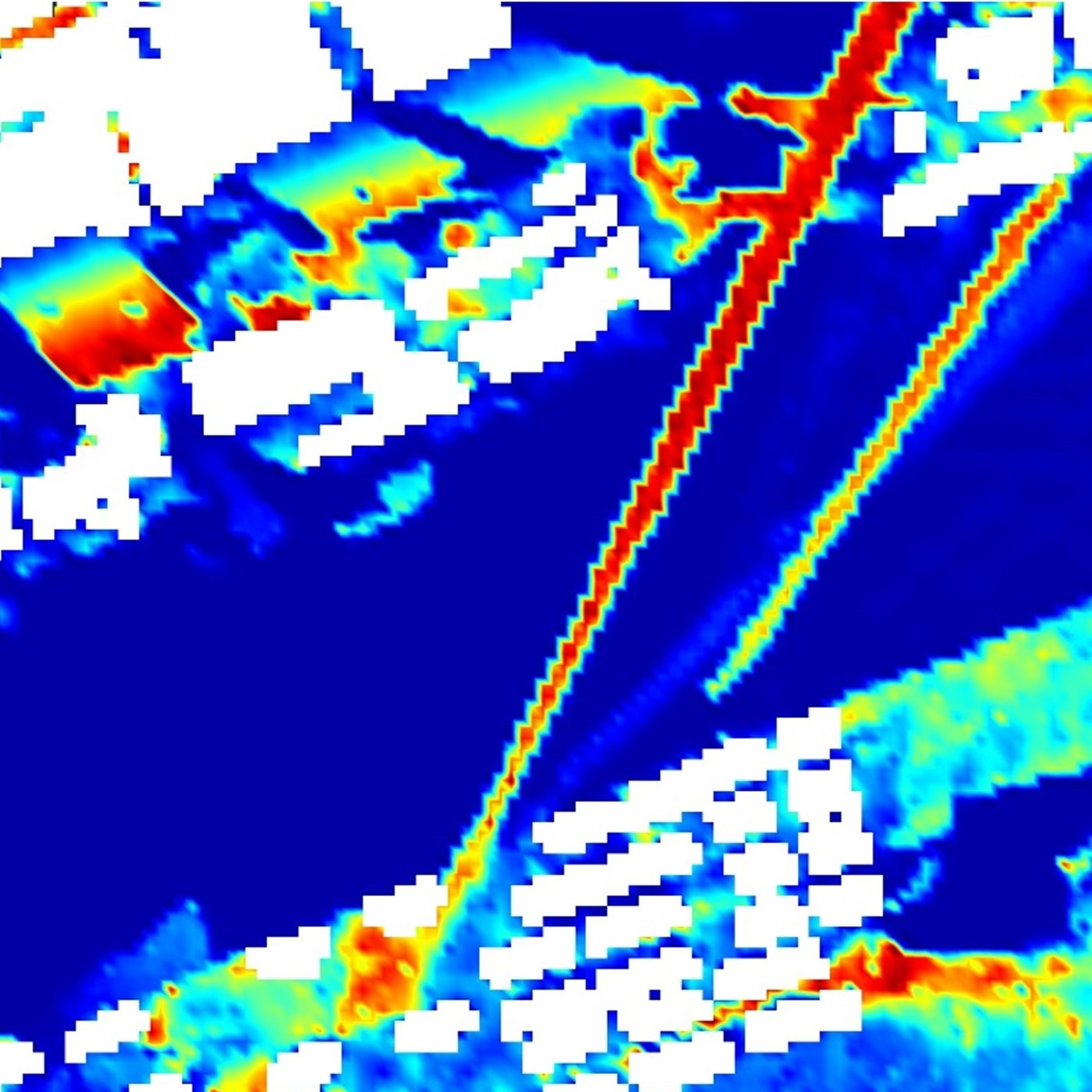}}
	\subfigure[GAN]{
		\label{GAN}
		\includegraphics[width=0.31\linewidth]{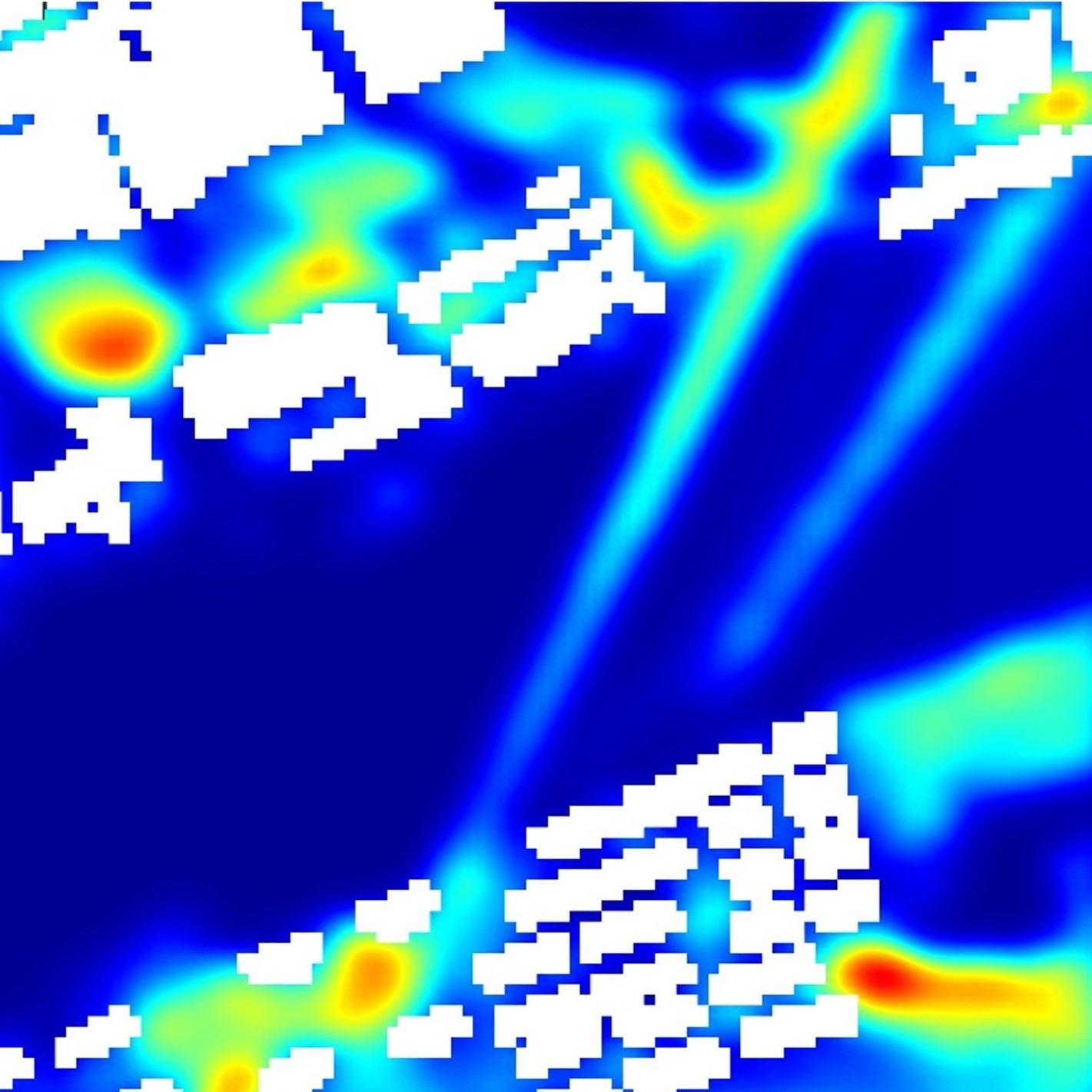}}
	\subfigure[ResNet50]{
		\label{resnet50}
		\includegraphics[width=0.31\linewidth]{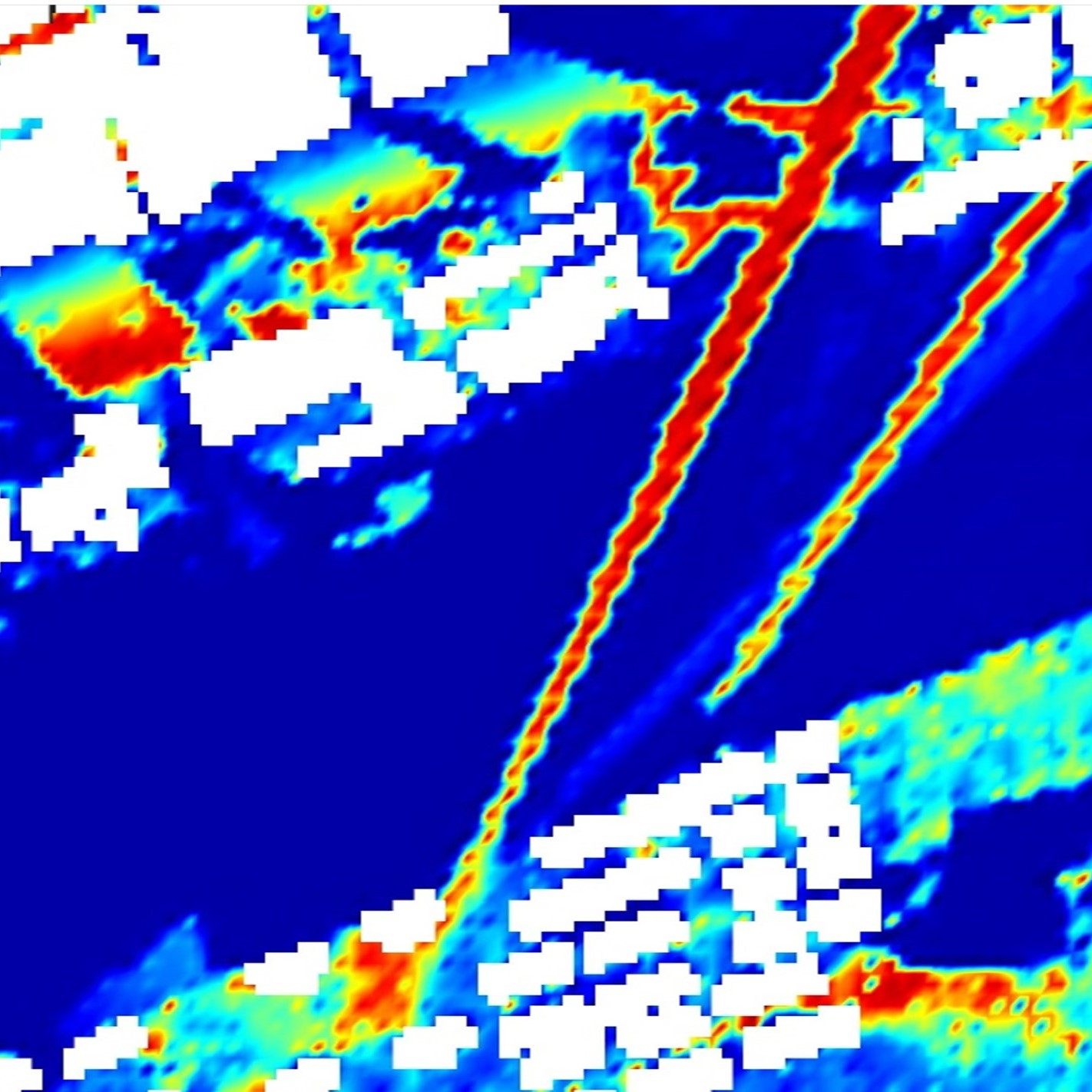}}
	\caption{Visualization of Super-Resolution results of DS in a region of 1000 × 1000 $m^2$ with scale factor 2.}
	\label{Visualization}
\end{figure}

It can be observed from the experiment results that, the MAE and absolute mean error(AME) of the proposed model on the PL achieve the effect of 2.82 dB and 0.3 dB with scale factor 2. Under the same conditions, the comparison results of ResNet50, Vit, GAN, and UNet are shown in Fig. \ref{Visualization} and Table \ref{net-compare}. Obviously, after testing, it can be found that these DL models, which achieved good results in other fields such as computer vision or natural language processing, performed worse than the proposed model in this paper. On the single-task fitting for PL, the performance of ResNet50 and UNet is around 7\--{}8 dB on average after several epochs of training. Moreover, ViT can only reach 8 dB after modifying more layers and processing with masks. The best result of GAN is even larger than 12 dB and still contains much noise. Our model has achieved results far exceeding the popular SR models on the channel super-resolution task through the comparison.

On the RT-Urban dataset, we performed SR training on 6 main channel features with scales of 2, 4, and 8, and the results are shown in Figure. \ref{trainprocess} and Table \ref{MAE error}. During the training process of the shared parameter layer, it can be found that the loss of each feature decreases rapidly around the first 100 epochs. The performance of the following targeted feature extractor can be optimized by 0.5\--{}0.7 dB on the best achieved by the backbone after fine-tuning.

In the classification training of LOS/NLOS and $\theta$, it can be observed that the second half of SR training at scale 8 is more volatile and not as smooth as other features. After splitting the feature map data and visual analysis, we found that large-scale downsampling will make the edge of the classification area seriously jagged. Moreover, extracting a more accurate mapping relationship is impossible and will affect accuracy. However, the classification can still achieve a correct rate of more than 91\%, indicating that the proposed model has an imposing recovery effect on the channel feature data.

\begin{table}[h]
\renewcommand\arraystretch{1.2}
\caption{Quantitative results for the experiments} 
\label{net-compare}
\begin{center}

\begin{tabular}{l|llll}
\hline
\multicolumn{2}{l|}{\textbf{Method}} & \textbf{AME/dB} & \multicolumn{1}{l}{\textbf{MAE/dB}} & \multicolumn{1}{l}{\textbf{RMSE}} \\ \hline
\multicolumn{2}{l|}{UNet}               & > 1              & > 7        & > 13                                                        \\
\multicolumn{2}{l|}{GAN}  & > 2               & > 12                 &  > 20                                               \\
\multicolumn{2}{l|}{GANSR}  & > 2               & > 12                  & >  17                                              \\
\multicolumn{2}{l|}{ResNet50}            & 0.5-1               & 5-6                 & 9-10                                                 \\
\multicolumn{2}{l|}{ViT}           & 0.8-1               & 6-7                    & 12-14                                             \\
\multicolumn{2}{l|}{\textbf{Ours}}         & \textbf{0.3(best)}               & \textbf{2.82}            &\textbf{5.11}                                              \\
\hline
\end{tabular}
\end{center}
\end{table}

\begin{table}[h]
\centering
\caption{Super resolution performance (AME \& RMSE) } \label{MAE error}
\resizebox{\linewidth}{!}{
\begin{tabular}{|c|c|c|c|c|c|c|c|} 
\hline
\multicolumn{2}{|c|}{\diagbox{Scale}{Targets}}            & PL                                     & $R_p$                                  & DS                                     & $\phi$                            & $\theta$                           & \begin{tabular}[c]{@{}c@{}}\scriptsize{LOS}/\\ \scriptsize{NLOS}\end{tabular}                              \\ 
\hline
\multirow{2}{*}{2} & AME                                  & 0.30                                      & 0.05                                      & 0.29                                      & 0.38                                      & 0.13                                      & 0.01                                     \\ 
\hhline{|~-------|}
                   & {\cellcolor[rgb]{0.89,0.89,0.89}}RMSE & {\cellcolor[rgb]{0.89,0.89,0.89}}5.11 & {\cellcolor[rgb]{0.89,0.89,0.89}}1.93  & {\cellcolor[rgb]{0.89,0.89,0.89}}10.64 & {\cellcolor[rgb]{0.89,0.89,0.89}}9.65 & {\cellcolor[rgb]{0.89,0.89,0.89}}1.60  & {\cellcolor[rgb]{0.89,0.89,0.89}}N/A  \\ 
\hline
\multirow{2}{*}{4} & AME                                  & 0.54                                      & 0.14                                      & 1.01                                      & 1.18                                      & 0.18                                      & 0.05                                      \\ 
\hhline{|~-------|}
                   & {\cellcolor[rgb]{0.89,0.89,0.89}}RMSE & {\cellcolor[rgb]{0.89,0.89,0.89}}7.09 & {\cellcolor[rgb]{0.89,0.89,0.89}}3.08 & {\cellcolor[rgb]{0.89,0.89,0.89}}15.97 & {\cellcolor[rgb]{0.89,0.89,0.89}}15.17 & {\cellcolor[rgb]{0.89,0.89,0.89}}2.24 & {\cellcolor[rgb]{0.89,0.89,0.89}}N/A  \\ 
\hline
\multirow{2}{*}{8} & AME                                  & 0.71                                      & 0.28                                      & 2.72                                      & 2.51                                      & 0.36                                      & 0.10                                      \\ 
\hhline{|~-------|}
                   & {\cellcolor[rgb]{0.89,0.89,0.89}}RMSE & {\cellcolor[rgb]{0.89,0.89,0.89}}8.99 & {\cellcolor[rgb]{0.89,0.89,0.89}}4.30 & {\cellcolor[rgb]{0.89,0.89,0.89}}20.71  & {\cellcolor[rgb]{0.89,0.89,0.89}}20.34 & {\cellcolor[rgb]{0.89,0.89,0.89}}2.97 & {\cellcolor[rgb]{0.89,0.89,0.89}}N/A  \\ 
\hline

\end{tabular}
}

\end{table}

\begin{table}[t]
\caption{Cumulative super-resolution performance of PL  }\label{ablationstudy}
\begin{center}
\begin{threeparttable} 
\resizebox{\linewidth}{!}{
\begin{tabular}{|l|c|c|c|c|c|c|} 
\hline
\multirow{2}{*}{} & \multicolumn{3}{c|}{\textbf{MAE}} & \multicolumn{3}{c|}{\textbf{RMSE}} \\ 
\cline{2-7}
 & scale=2 & scale=4 & scale=8 & scale=2 & scale=4 & scale=8 \\ 
\hline
+ATT & +5\% & +3\% & +6\% & +6\% & +4\% & +5\% \\ 
\hline
+DA & +16\% & +12\% & +12\% & +10\% & +5\% & +6\% \\
\hline
+RES & +11\% & +7\% & +9\% & +11\% & +5\% & +6\% \\ 
\hline
STL & 0 & 0 & 0 & 0 & 0 & 0 \\ 
\hline
\end{tabular}
}
\begin{tablenotes}    %这行要添加， 从这开始
        \footnotesize           %这行要添加
        \item\leftline{+ATT: Add attention mechanism to +DA.} 
        \leftline{+DA: Add data augmentation to +RES.}
        \leftline{+RES: Add residual connection and \textit{iterative up-and-down} to STL} 
        \leftline{STL: The proposed model without RES, DA, and ATT in training.} 
        % \end{footnotesize}
      \end{tablenotes}            %这行要添加

\end{threeparttable}
\end{center}
\end{table}
\subsection{Ablation study}
The ablation experiments are used to verify whether these methods we take to improve the SR of the model improve the fitting effect. After removing different strategies, we choose the most representative PL among the channel characteristics and test its performance with MAE and RMSE. The results of the MAE and RMSE ablation experiments are shown in Table \ref{ablationstudy}. Here we use the proposed model as the baseline. RES represents the strategy of residual connections in the neural network, and the attention mechanism will improve the super-resolution results by nearly 23\%. DA will also have a significant effect on MAE. And it can be noticed that even with the increase of the SR scale, the improvement by these methods will still be stable.

\section{Conclusion}
This paper proposes a residual-based SR model for wireless channel characteristics. We enhance the fitting ability of the proposed SR model by attention mechanism and generate a higher accuracy. A deep-shallow panel is used to expand the receptive field. We train our model using RT-urban constructed by CloudRT platform. The proposed model can achieve SR performances of PL with MAE of 2.83 dB and 99\% accuracy of LOS areas given scale factor of 2. As the SR scale increases, this model maintains stable performance according to the numerical experiments. The proposed model is also compared with other state-of-the-art DL models such as ResNet, ViT, and GAN. In the future, we study the structure used in our current SR model to improve the accuracy of the SR of channel characteristics.
\section*{Acknowledgements}
This work is supported by the Fundamental Research Funds for the Central Universities2022JBXT001, the Ministry of Education of China under Grant 8091B032123, NSFC under Grant 62271043, and Beijing Natural Science Foundation L221009.
\bibliographystyle{unsrt}
\bibliography{ref}

\end{document}